 \documentclass[sigconf]{acmart}

\AtBeginDocument{%
  \providecommand\BibTeX{{%
    \normalfont B\kern-0.5em{\scshape i\kern-0.25em b}\kern-0.8em\TeX}}}

\copyrightyear{2020}
\acmYear{2020}
\setcopyright{acmcopyright}\acmConference[CHI PLAY '20 EA]{Extended Abstracts of
the 2020 Annual Symposium on Computer-Human Interaction in Play}{November
2--4, 2020}{Virtual Event, Canada}
\acmBooktitle{Extended Abstracts of the 2020 Annual Symposium on
Computer-Human Interaction in Play (CHI PLAY '20 EA), November 2--4, 2020,
Virtual Event, Canada}
\acmPrice{15.00}
\acmDOI{10.1145/3383668.3419958}
\acmISBN{978-1-4503-7587-0/20/11}

\begin{document}

\title{VirusBoxing: A HIIT-based VR boxing game}

\author{Wenge Xu}
\affiliation{%
  \institution{Xi'an Jiaotong-Liverpool University}
  \city{Suzhou}
  \state{Jiangsu}
  \country{China}}

\author{Hai-Ning Liang}
\authornote{Corresponding author: haining.liang@xjtlu.edu.cn}
\affiliation{%
  \institution{Xi'an Jiaotong-Liverpool University}
  \city{Suzhou}
  \state{Jiangsu}
  \country{China}
}

\author{Xiaoyue Ma}
\affiliation{%
  \institution{Xi'an Jiaotong-Liverpool University}
  \city{Suzhou}
  \state{Jiangsu}
  \country{China}}

\author{Xiang Li}
\affiliation{%
  \institution{Xi'an Jiaotong-Liverpool University}
  \city{Suzhou}
  \state{Jiangsu}
  \country{China}}

\renewcommand{\shortauthors}{Xu et al.}

\begin{abstract}
  Physical activity or exercise can improve people's health and reduce their risk of developing several diseases; most importantly, regular activity can improve the quality of life. However, lack of time is one of the major barriers for people doing exercise. High-intensity interval training (HIIT) can reduce the time required for a healthy exercise regime but also bring similar benefits of regular exercise. We present a boxing-based VR exergame called VirusBoxing to promote physical activity for players. VirusBoxing provides players with a platform for HIIT and empowers them with additional abilities to jab a distant object without the need to aim at it precisely. In this paper, we discuss how we adapted the HIIT protocol and gameplay features to empower players in a VR exergame to give players an efficient, effective, and enjoyable exercise experience.
\end{abstract}




\maketitle

\section{Introduction}
Physical inactivity has been identified as the fourth leading cause of death globally (6\% of deaths globally), while being overweight and obesity are responsible for 5\% of global mortality \cite{WHO_2009}. These factors are responsible for raising the risk of heart disease, diabetes, and cancers across countries in all income groups: high, middle, and low \cite{WHO_2009}. It has been shown that participation in regular physical activity could not only reduces the risk of various diseases, stroke, diabetes, but is also fundamental to energy balance and weight control \cite{WHO_Global_2010}. Despite the known benefits of regular moderate-intensity exercise in regulating risk factors leading to diseases, most people are still physically inactive \cite{american_college_of_sports_medicine_acsms_2010}. One of the most commonly cited barriers for people to exercise is lack of time \cite{barathi_interactive_2018}.

High-intensity interval training (HIIT) consists of alternating short bursts of intensive aerobic exercises with periods of passive or active low-intensity exercises \cite{Fox1973IntensityAD}. These combinations could reduce the time required for a healthy exercise regime. Studies show that HIIT is equally beneficial or superior to traditional aerobic exercise in many fitness and health-related measures \cite{Norris1998EffectsOE,gibala_metabolic_2008,laursen_scientific_2002}. Moreover, it offers the possibility to maintain high-intensity exercise for far more extended periods than doing continuous exercises of the same type \cite{billat_interval_2001}. Therefore, employing HIIT as a training protocol could be useful, especially for people who have limited time. 

In recent years, exergames, which combines video gaming and exercise, have been proposed as a solution to improve motivation and long-term engagement of people with doing regular exercises \cite{siegel_active_2009}. Previous literature has shown that exergames could bring physical and mental health outcomes to players of different age groups (e.g., older adults \cite{xu_health_benefits_2020}). The benefits of playing exergames include, but are not limited to, improved quality of life \cite{da_silva_alves_effects_2018}, acute cognitive capabilities \cite{gao_acute_2012}, and balance \cite{balance_2018}.

Virtual reality (VR) exergames \cite{xu_assessing_2019,xu_results_2020} with head-mounted displays (HMDs) have been gaining rapid attention. For instance, Xu et al. \cite{Xu_Studying_VR} found that playing exergames in VR was more challenging, immersive (with respect to flow, sensory and imaginative immersion), and had a lower negative effect than in front of a 50-inch TV. Studies have also suggested that playing exergames in VR helps promote physical activity in sedentary and obese children \cite{rizzo_virtual_2011}, especially to increase their motivation to exercise \cite{plante_does_2003,mestre_does_2011}. Moreover, VR can give users the illusion of grater capabilities (e.g., running and jumping), which can increase intrinsic motivation, perceived competence and flow, and may also increase motivation for physical activity in general \cite{ioannou_virtual_2019}.

\begin{figure*}[h]
  \centering
  \includegraphics[width=\textwidth]{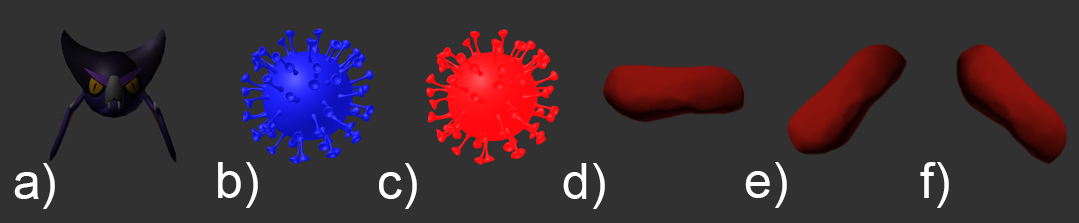}
  \caption{a) Devil virus creator that is placed 15meters away from the player. b) Blue virus that can be destroyed by a left-hand jab. c) Red virus that can be destroyed by a right-hand jab. d) Flat blood cell that requires the player to avoid hitting by squatting. e) Right tilted blood cell that requires the player to avoid hitting by squatting and leaning the body towards the right. f) Left tilted blood cell that requires the player to avoid hitting by squatting and leaning the body towards the Left.} 
  \Description{models used in the game}
\end{figure*}

We have developed a game called VirusBoxing that combines boxing with HIIT and use VR to enable players to have additional powers. Boxing is chosen as our exercise activity because it could be feasible and effective with a HIIT protocol \cite{cheema_feasibility_2015}. During 7 minutes of gameplay with VirusBoxing, players are required to jab the virus and weave the blood cell. The player could earn one energy unit when they jab a virus in the non-empowerment condition. Once they hit 10 viruses (i.e., when the energy bar is full), they could choose to activate the augmented ability for 10 seconds.

\section{Related Work}
Employing a HIIT protocol in exergame is not new. Research suggested that employing a HIIT protocol could significantly increase the energy expenditure and heart rate when compared to a non-HIIT version \cite{Nickel_2012}. Moreover, stationary cycling-based VR exergames that employed HIIT have been found to let players achieve the required intensity for HIIT \cite{barathi_interactive_2018,keesing_hiit_2019,haller_hiit_2019}. Applying HIIT to VR exergame could also potentially mitigate the typical VR HMD-related issues such as VR sickness \cite{Xu_Studying_VR}, sweat and wearer discomfort \cite{shaw_challenges_2015}.

There is a limited number of studies that have focused on ability empowerment in VR. Granqvist et al. \cite{granq_2018_chiplay} explored the empowered flexibility of the avatar in a martial arts VR game and found that a medium degree of empowered flexibility was preferred over realism or strong exaggeration of users' power. Ioannou et al. \cite{ioannou_virtual_2019} explored a jump- and running-in-place application with applied forward motion in a VR exergame and found increased immersion and motivation of players when they played with their ability empowered. Overall, providing ability empowerment to players could be useful in exergames.

In commercial games, BoxVR is the closest to our work. Although the game mechanics are similar (i.e., jabbing and weaving), our game is different in the following aspects: (1) Training protocol: BoxVR relies on a rhythm-based training where the intensity of the exercise may not be sufficient enough, but VirusBoxing employed a HIIT approach, which as shown in the literature we reviewed earlier and also our initial results (presented later in the paper), can be effective. (2) VirusBoxing maximizes the potential of VR to provide users ability empowerments in punching (i.e., players can punch a distant object without precisely aiming at the target). (3) VirusBoxing does not force players to be in the empowered ability mode all the time; instead, it allows the player to switch on the empowerment mode when they have enough energy.

\section{Overall Concept and Mechanics}
VirusBoxing combines boxing-inspired workouts and HIIT protocol for VR HMDs. In VirusBoxing, the player becomes the virus fighter who needs to protect the world behind himself/herself from a virus creator that is placed 15 meters away (see Fig. 1a). The virus creator could spawn viruses (see Fig. 1b-c) and blood cells (see Fig. 1d-f) towards the player. 

The player needs to jab the virus at a certain speed (i.e., at least 1m/s in our game). There are two colors of the virus: blue viruses can only be destroyed by a left-hand jab while red viruses can only be destroyed by a right-hand jab. Meanwhile, the player needs to weave blood cells in three ways (see Fig. 1d-f), either by (1) squatting only, (2) squatting and leaning the body rightwards, or (3) squatting and leaning the body leftwards.

The player earns one energy unit every time it successfully jabs on a virus, and once the energy reaches the maximum number of 10 units, the player could choose to switch on the ability empowerment mode that lass for 10 seconds, during which they could hit a distant virus without precisely targeting at it.

The percentage of spawning a virus and a blood cell is 70\% and 30\%, respectively. The percentage of spawning a red virus and a blue virus are the same (i.e., 35\%). For a flat blood cell (see Fig. 1d) is 20\% and 5\% for right tilted blood cell (see Fig. 1e) and left tilted blood cell (see Fig. 1f). 

\section{Innovation}
Our game includes three unique selling points when compared to other games in the market. First, it includes a HIIT-based workout. Second, it includes features of ability empowerment in how punches can be performed. Third, we have invented an energy system to activate the ability empowerment. Details of each innovation will be described in the following subsections.

\subsection{High Intensity Interval Training Protocol}
Unlike BoxVR which implies a rhythm-based approach for a VR boxing experience, our exergame employed a low volume HIIT protocol, which has the following advantages: (1) HIIT has been proved to be effective and efficient to gain health benefits (e.g., increase one's exercise capacity \cite{jacob_hiit}) while the benefits of a rhythm-based approach for people's health are unknown; (2) the HIIT protocol suits VR particularly well since a short gameplay duration mitigates typical HMD usability problems such as VR sickness \cite{Xu_Studying_VR}, sweat and wearer discomfort \cite{shaw_challenges_2015}; (3) it is essential for an exercise routine to include both oxygen-dependent (aerobic) and oxygen-independent (anaerobic) exercises \cite{rozenek_physiological_2007, billat_interval_2001}.

We employed the following protocol: 3 sessions of 30-second low intensity and 90-second sprint, ending with a 60-second low intensity cool down. In the low-intensity interval, the game would spawn a game object every 0.8s and the moving speed of the game object is 5.7 m/s. In the high-intensity interval, the game would spawn a game object every 0.5s and the moving speed of the game object is 8 m/s. Players need to put as much effort as possible (e.g., punch as heavy as they can). 

\subsection{Punching Empowerment in VR}
\subsubsection{Study Design}
We took into account two factors when we designed an empowered punching experience in VR boxing games. The first one was the punching range, which had three levels: 1) short-range (RA) --- up to 5m, 2) medium-range (MR) --- up to 10m, and 3) long-range (LR) --- up to 15m. The second factor was the need for precise targeting, which had two levels: 1) precise targeting (PT) --- the player needs to target the virus precisely when jabbing, and 2) rough targeting (RT) --- the player only need to punch with the correct hand.

We did not test both factors in one study at one time. Instead, we have set up two experiments for these factors with the same group of participants on different days. The reasons are 1) we want to test the factors that might affect the ability empowerment of punching gradually, 2) the length of the experiment might be too long: two-way within-subjects design of this study would ask participants to do VR HIIT-based boxing for 42 minutes (6 conditions $\times$ 7 minutes gameplay of each condition) excluding the time that required for participants to fill the questionnaire and rest. 

\begin{table}
  \caption{Summary of the results of the punching range study according to game experience and simulator sickness.}
  \label{tab:freq}
  \begin{tabular}{p{1.58cm} p{1.53cm} p{1.53cm}p{1.53cm} p{0.5cm}}
    \toprule
   Component& SR & MR & LR & p\\
    \midrule
    Competence & 11.17 (5.36) & 11.33 (4.89) & 10.25 (4.60) & $p$=0.508\\
     Tension & 1.75 (2.49) & 0.92 (1.68) & 1.17 (1.85) & $p$=0.535\\ 
     Sensory and Imaginative Immersion & 10.33 (7.30) & 10.08 (7.44) & 10.17 (7.46) & $p$=0.938\\
    Flow & 9.83 (6.07) & 9.58 (6.57)& 8.92 (5.518) & $p$=0.683\\
    Negative Affect & 4.08 (3.45) & 3.00 (2.05)& 4.17 (3.04) & $p$=0.470\\
    Positive Affect & 10.00 (6.38) & 10.08 (6.07)& 9.25 (5.01) & $p$=0.544\\
    Challenge & 5.83 (4.47) & 5.83 (5.46)& 6.00 (5.53) & $p$=0.987\\
    Nausea & 10.34 (14.92) & 13.52 (21.32)& 13.52 (18.84) & $p$=0.596\\
    Oculomotor & 20.85 (26.28) & 20.21 (21.52)& 22.74 (22.62) & $p$=0.886\\
    Disorientation & 23.20 (32.15) & 17.40 (23.83)& 17.40 (26.63) & $p$=0.427\\
  \bottomrule
\end{tabular}
\end{table}

\begin{table}
  \caption{Summary of the results of targeting conditions according to game experience and simulator sickness. * symbol indicates there is a significant difference.}
  \label{tab:freq}
  \begin{tabular}{p{2.4cm} p{1.8cm} p{1.8cm} p{0.9cm}}
    \toprule
   Component& PT & RT & p\\
    \midrule
    Competence & 10.75 (5.67) & 11.5 (5.72) & $p$=0.351\\
     Tension & 1.42 (1.62) & 0.67 (0.89) & $p$=0.032*\\ 
     Sensory and Imaginative Immersion & 9.08 (8.58) & 10.42 (7.77) & $p$=0.104\\
    Flow & 9.33 (5.63) & 9.33 (6.30) & $p$=1.000\\
    Negative Affect & 4.42 (2.57) & 5.75 (4.22) & $p$=0.352\\
    Positive Affect & 8.83 (6.63) & 10.33 (6.21) & $p$=0.204\\
    Challenge & 5.33 (4.29) & 4.83 (3.49) & $p$=0.429\\
    Nausea & 11.93 (17.79) & 11.93 (17.79) & $p$=1.000\\
    Oculomotor & 18.32 (16.31) & 15.79 (16.94) & $p$=0.438\\
    Disorientation & 20.88 (27.52) & 19.72 (28.75) & $p$=0.820\\
    
  \bottomrule
\end{tabular}
\end{table}

\subsubsection{Participants, Apparatus, and Procedure}
We recruited 12 participants (3 female, 9 male; age 19-21, mean 19.75; 7 with previous VR experience, but only 1 was regular user and none of them have experience with the device used in our experiment [i.e., Oculus Quest]) to join our experiments 1 and 2 to inform our design choices. Participants were screened using the Physical Activity Readiness Questionnaire (PAR-Q) \cite{thomas_revision_1993}. If a participant answered 'yes' to any of the PAR-Q questions, they were excluded from doing the experiment.

Participants were then asked to complete pre-experiment questionnaires to collect their demographics information and were given information about the study and asked to fill the  consent form. Participants then had a chance to become familiar with the device since they were all new to the Oculus Quest. When participants felt rested and ready, they would proceed to the testing session (i.e., playing the 7-mins version of VirusBoxing). Conditions were counterbalanced across participants. After each condition, participants needed to complete game experience questionnaire \cite{ijsselsteijn_game_2008} and simulator sickness questionnaire \cite{kennedy_simulator_1993}. At the end of each experiment, we asked them to provide feedback on the virus spawn frequency and flying speed, distribution of each gameobject, and ranked the game versions.

\begin{figure*}[t]
  \centering
  \includegraphics[width=\textwidth]{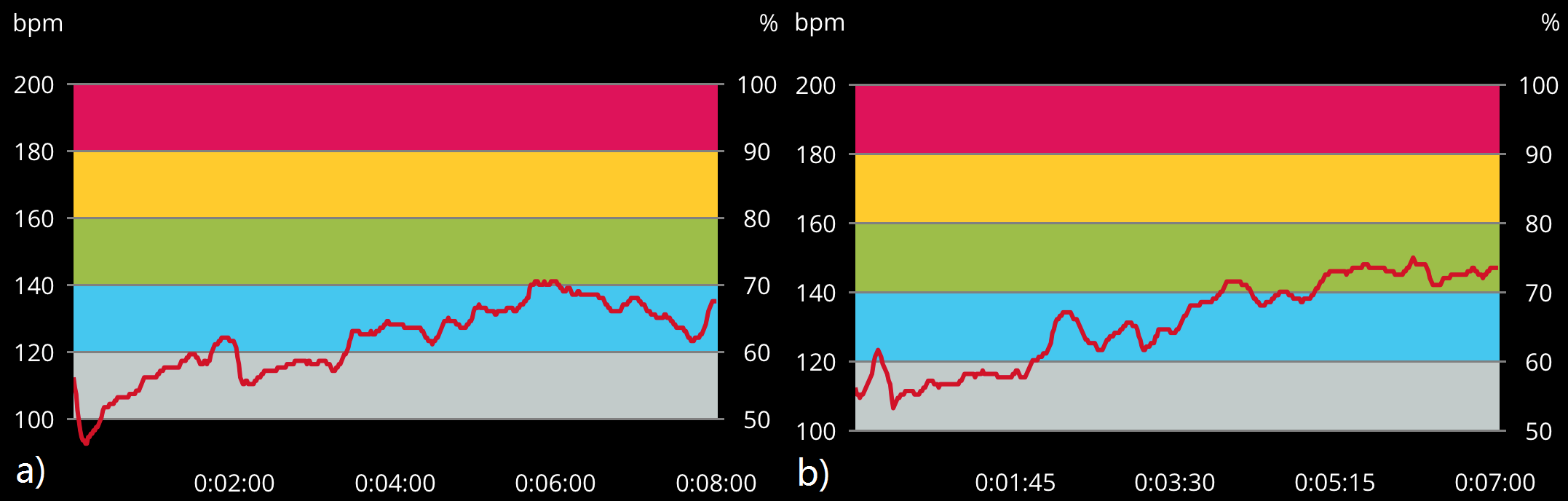}
  \caption{Participant's heart rate throughout the 7 minutes of gameplay: (a;left) the participant who exercise regularly, (b;right) the participant who does not exercise regularly.}
  \Description{Real time HR}
\end{figure*}

\subsubsection{Punching Range Study Results and Discussion}
Gameplay performance data were analyzed in the percentage of missing viruses and the percentage of hitting blood cells during the game.
The percentage of the missing virus was 7.03\% (SD=7.08\%) for SR, 15.80\% (SD=16.95\%) for MR, and 8.76\% (SD=8.17\%) for LR. One-way Repeated ANOVA tests showed that there were no significant differences between the three versions ($F_{2,22}=2.800, p=.083$). The percentage of hitting blood cells were 0.00\% (SD=0.00\%) for all conditions, which means all players could avoid hitting all blood cells when playing the game. We did not found any significant difference between all three versions regarding game experience (Competence, Tension, Sensory and Imaginative Immersion, Flow, Negative Affect, Positive Affect, Challenge) and simulator sickness (nausea, oculomotor, disorientation) (see Table 1). Regarding the ranking, 8 participants preferred the LR version.

We decided to apply the LR as the punching range for the punching empowerment since the majority of participants preferred the LR. Also, participants argued that it was difficult to predict whether the virus has reached their punching range in SR and MR versions.

\subsubsection{Targeting Condition Study Results and Discussion}
In this section, we analyze data by using pair-sample t-tests. There was a significant effect of targeting condition ($t(11)=2.853, p=.016$) on the percentage of missing viruses, showing that players missed more viruses in PT (M=18.98\%, SD=15.24\%) than RT (M=7.12\%, SD=5.57\%). Regarding game experience and simulator sickness, paired-sample t-tests yielded a significant difference of Tension among conditions ($t(11)=2.462, p=.032$) where RT caused significantly lower Tension than PT. No other significant effects were observed on game experience and simulator sickness (see Table 2). There was no clear preference for targeting conditions from the ranking data (i.e., each version was preferred by 6 participants).

Because players felt a lower tension (annoyed, frustrated, irritable) and had a higher gameplay performance in the RT during the gameplay, we employed RT. In conclusion, the empowered punches could hit a virus within 15 meters and would not require the player to  target it in a precise manner. 

\subsection{Energy System}
Unlike the previous study where empowerment is given throughout the game experience \cite{ioannou_virtual_2019}, empowerment in VirusBoxing can only be activated when following conditions are met: 1) the energy system is full and 2) players decide to switch on empowerment by pressing the button 'A' on the right controller. In order to gain energy for empowerment, players need to jab the virus in the non-empowerment condition. 

\section{Playtesting}
A short playtesting was conducted with 2 players. The focus here was mainly on exertion. We employed Polar OH1 to capture the heart rate of the player. One player (aged 19; male; regular gym visitor --- 3 or 4 times per week; body mass index: 21.1) experienced a max heart rate of 140 beats per minute (bpm), an average of 122 bpm and burned 44kcal calories across the whole session (see Fig. 2a). The other player (aged 19; male; lack of exercise; body mass index: 27.4) experienced a max heart rate of 149 bpm, an average of 130 bpm and burned 44kcal calories across the whole session (see Fig. 2b). However, we could not have further participants due to issues brought in by the COVID-19 pandemic. Overall, the experience seems to be effective and both players highly recommended VirusBoxing (rated the experience to be 4 out of 5).

\section{Target Audience}
It is now well established that a sedentary lifestyle is a unique risk factor for illnesses such as Type 2 Diabetes and cardiovascular disease \cite{williams_obesity:_2009}, which account for 30\% of global mortality. Our game is designed to help tackle this problem and provide our target audience (i.e., people with a sedentary lifestyle) with a platform to let players have short but meaningful and fun exercise experiences. 

\section{Future Work}
Future work could focus on conducting an experiment with our game regarding exertion with more participants and exploring the use of PID control to maintain the player's heart rate at a high level in a high-intensity sprint period and help to lower the heart rate in the recovery period.

\section{Conclusion}
This paper introduces VirusBoxing as a promising intervention of exercise for all age groups. VirusBoxing's design combines three concepts to achieve a fun, effective, exercise experience: (1) A high-intensity interval training protocol which incorporates the benefits of exercise, (2) ability empowerment to give players the illusion of greater capabilities than they actually have (i.e., punching a distant object), and (3) an energy system which enriches the traditional boxing-based VR games.

\begin{acks}
The authors would like to thank the anonymous reviewers for their valuable comments and helpful suggestions. The work is supported in part by Xi’an Jiaotong-Liverpool University (XJTLU) Key Special Fund (\#KSF-A-03) and XJTLU Research Development Fund.
\end{acks}



\end{document}